\documentclass[article,11pt]{IEEEtran}

\usepackage[english]{babel}
\usepackage[utf8]{inputenc}
\usepackage{amsmath}
\usepackage{graphicx}
\usepackage[T1]{fontenc}
\usepackage[utf8]{inputenc}
\usepackage{authblk}
\usepackage{cite}
\usepackage{caption}
\usepackage{enumerate}
\usepackage[colorinlistoftodos]{todonotes}
\title{\vspace{-1.0cm}CCNCheck: Enabling Checkpointed Distributed Applications in Content Centric Networks}

\author{\large Nitinder Mohan}
\author{\large Pushpendra Singh}
\affil{Indraprastha Institute of Information Technology (IIIT), New Delhi, India}

\date{\today}

\begin{document}
\maketitle

\begin{abstract}
We consider the problem of checkpointing a distributed application efficiently in Content Centric Networks so that it can withstand transient failures. We present CCNCheck, a system which enables a sender optimized way of checkpointing distributed applications in CCN's and provides an efficient mechanism for failure recovery in such applications. CCNCheck's checkpointing mechanism is a fork of DMTCP repository   CCNCheck is capable of running any distributed application written in C/C++ language.  
\end{abstract}

\vspace{-0.4cm}
\section*{\textbf{Motivation}}

As CCN offers receiver-driven mode of communication, the distributed applications running on it needs to be modified from their usual sender-driven paradigm\cite{voccn}. Checkpointing and rollback-recovery are well known techniques that allow processes to make progress in spite of failures\cite{koocheckpoint}. However, CCN is devoid of any such mechanism of failure recovery. Keeping the above points in mind, we bring CCNCheck which offers a sender-optimized way of running distributed applications. CCNCheck also implements checkpointing for applications running on CCN.

\vspace{-0.3cm}
\section*{\textbf{Our Contribution}}
A typical distributed application in CCN is assumed to be running on multiple nodes and uses a common channel to send/receive interests and data. We further assume that: 
\begin{enumerate}
\item Processes do not have any common clock/ memory.
\item Processes follow a fail-stop model of failing i.e. processes can crash by stopping execution and remain halted until restarted. 
\end{enumerate}
Checkpoint is saved local state of a process. Set of local states and messages in common channel is global state of a system. In checkpointing, every process takes a local checkpoint to ensure a global consistent state which can later be used to recover a system from failure\cite{chandysnapshots}.\\
Our work is centered around using interests as notifications/signals in CCN. Formally, a distributed system running on CCNCheck works on following model: 
\begin{enumerate}
\item Every node knows about other nodes running the same distributed application.
\item Every node is defined by a unique name which is pre-appended by the application name the process is running.
\item The interest packet is not stored in the router's cache. The router only forwards the interest using its FIB entry.
\end{enumerate}
\subsection{Running Distributed Application}
\begin{figure}
\centering
\captionsetup{justification=centering}
\includegraphics[width=0.3\textwidth]{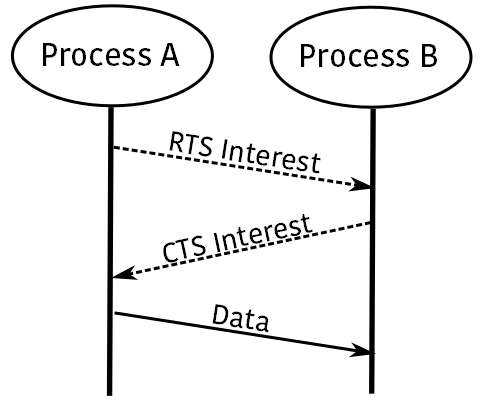}
\caption{\label{fig:1}CCNCheck Communication Handler}
\end{figure}

To run a sender-driven distributed application in CCN we use an approach very similar to solving hidden terminal problem in IEEE 802.11 networks. The sender first issues a \textbf{Request-to-Send (RTS)} interest to the desired destination process. This RTS packet acts as a notification to destination about an incoming data. The name of RTS contains the identifying name of its issuer using which the destination process issues a \textbf{Clear-to-Send (CTS)} interest back to the sender. CTS serves as the necessary interest required to send the data in CCN. Figure \ref{fig:1} depicts process A sending some data to process B using CCNCheck.

\subsection{Enabling Checkpoint}
Distributed Multi-Threaded Checkpoint (DMTCP) is a research-based, transparent, user-level checkpointing tool for distributed applications. DMTCP follows a blocking type algorithm of checkpointing to ensure a global consistent state at each checkpoint. It employs a stateless centralized coordinator to coordinate checkpoint requests between nodes\cite{dmtcp}.\\
We have developed a plugin for DMTCP which enables it to work in CCN environment. Even though CCN is deployed as an overlay on TCP/IP networks \cite{ccn2009} for which DMTCP works well, however, some more logical changes are necessary to make DMTCP function in CCN. We also formalize various inconsistent checkpoint scenarios due to uncoordinated checkpointing in CCN and devise a method to overcome such situations. Some of the changes made are:
\begin{enumerate}
\item DMTCP uses flush token to clear out TCP sockets during checkpoint process to ensure consistent checkpoint. As CCN works atop of Interests and Data packets, we have designed a "Flush Interest" which ensures that checkpoint is consistent from any orphan interests and data. 
\item DMTCP coordinator is modified to detect a CCN network and register itself with CCN Daemon on invoking. 
\item The Coordinator is run as a stateless process with a name unique to the environment/organization. We have designed interest packets which is used by coordinator to checkpoint processes in the application.
\item The restart from checkpoint process is able to resolve any non-responded interests due to lack of Pending Interest Table (PIT) entries. 
\item The discovery services in reconnect phase on restart from a checkpoint works using CCN namespaces. 
\end{enumerate}

\section*{\textbf{Implementation}}
\begin{figure}
\centering
\captionsetup{justification=centering}
\includegraphics[width=0.35\textwidth]{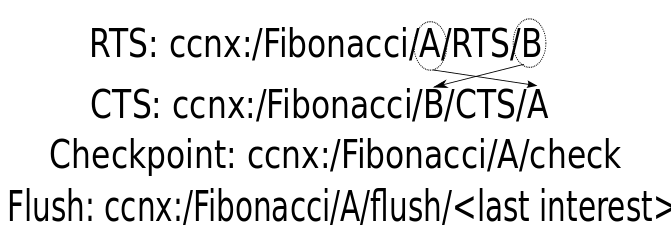}
\caption{\label{fig:2}Naming Rules in CCNCheck}
\end{figure}

\begin{enumerate}[I]
\item \textbf{\large System Model}\\
CCNCheck uses three layer abstraction model.
\begin{enumerate}
\item \textbf{Communication Handler:}  It handles the Interest and Data packets to be sent between communicating nodes. It is built on CCNx v0.8.2.
\item \textbf{Checkpoint Handler:} It provides the checkpoint mechanism in CCN and is based on DMTCP.  
\item \textbf{End-User Applications:} These are applications to be run in a distributed environment. It can be in C/C++ language.\\
\end{enumerate}
\item \textbf{\large Interest Naming Rules}\\
The naming format for RTS and CTS packets in CCNx are as follows:
\begin{center}
ccnx://Application Name/Receiver Address/Type of Interest/Sender Address
\end{center}
The \textit{Request-To-Send} and \textit{Clear-To-Send} interests use signal name 'RTS' and 'CTS' respectively. The checkpoint interest, however, is only one-way notification (i.e. from coordinator to process). Thus, interest name does not have the sender's name appended to it and it is denoted by signal type 'check'. Similarly, \textit{Flush Interest} has a signal name 'flush' but is appended with the last name of last interest sent  Figure \ref{fig:2} shows naming rules for CCNCheck.\\
\item \textbf{\large Applications}\\
CCNCheck was deployed on a test-bed of six interconnected nodes in CCN network. We have developed two sample distributed applications to review our system. We have also used an existing application to check the compatibility of our system.
\begin{enumerate}
\item A simple C application which keeps counting till infinity is run locally on each node with different start times and is killed later. The goal was to check the consistency of checkpoint taken by CCNCheck before failure.
\item A distributed C++ application in which the participating nodes compute the consecutive numbers of fibonacci sequence in an iterative manner. This application utilizes the distributed capabilities of CCNCheck to send the result to the next node after each subsequent computation.
\item A CCN enabled VLC player which can stream videos on a Content Centric Network.
\end{enumerate}
We are able to checkpoint all the applications listed above.
\end{enumerate}

\vspace{-0.4cm}
{\footnotesize \bibliographystyle{IEEEtran}
\small{\bibliography{references}}}

\end{document}